\documentclass[aps,longbibliography,
  twocolumn,amsmath,
  amssymb,nofootinbib, superscriptaddress, preprintnumbers]{revtex4-1}

\usepackage{graphics}      % standard graphics specifications
\usepackage{graphicx}      % alternative graphics specifications
\usepackage{longtable}     % helps with long table options
\usepackage{url}           % for on-line citations
\usepackage{bm}            % special 'bold-math' package
\usepackage{xcolor}
\usepackage{mathrsfs}
\usepackage[colorlinks,allcolors=blue]{hyperref}
\usepackage{blindtext}
\usepackage{hyperref}
\usepackage{float}
\usepackage{bbold}
\usepackage{lipsum}
\usepackage[normalem]{ulem}
\begin{document}
\preprint{MPP-2022-59}

\title{Oscillations of High-Energy Cosmic Neutrinos
in the Copious MeV Neutrino Background}
%: the Possibility of the Occurrence of Fast Neutrino Flavor Conversion Modes}

\newcommand\barparen[1]{\overset{(-)}{#1}}
\newcommand*{\MPP}{\textit{\small{Max-Planck-Institut f\"ur Physik (Werner-Heisenberg-Institut), F\"ohringer Ring 6, 80805 M\"unchen, Germany}}}
\newcommand*{\PSU}{\textit{\small{Department of Physics, Pennsylvania State University}}}
\author{Sajad Abbar}%Universit\'e Denis Diderot, 75205 Paris Cedex 13, France}
\affiliation{\MPP}
\author{Jose Alonso Carpio}
\affiliation{Department of Physics; Department of Astronomy and Astrophysics; Center for Multimessenger Astrophysics, Institute for Gravitation and the Cosmos, The Pennsylvania State University, University Park, Pennsylvania 16802, USA}
\author{Kohta Murase}
\affiliation{Department of Physics; Department of Astronomy and Astrophysics; Center for Multimessenger Astrophysics, Institute for Gravitation and the Cosmos, The Pennsylvania State University, University Park, Pennsylvania 16802, USA}
\affiliation{Center for Gravitational Physics, Yukawa Institute for Theoretical Physics, Kyoto, Kyoto 606-8502, Japan}

%\date{\today}

\begin{abstract}
The core-collapse of massive stars and merger of neutron star binaries are among the most promising candidate sites for the production of high-energy cosmic neutrinos. We demonstrate that the high-energy neutrinos produced in such extreme environments can experience efficient flavor conversions on scales much shorter than those expected in vacuum, due to their coherent forward scatterings with the bath of decohered low-energy neutrinos emitted from the central engine.
These low-energy neutrinos, which exist as mass eigenstates, provide a very special and peculiar dominant background for the propagation of the high-energy ones.
We point out that the high-energy neutrino flavor ratio is modified to a value independent of neutrinos energies, which is distinct from the conventional prediction with the matter effect. 
We also suggest that the signals can be used as a novel probe of new neutrino interactions beyond the Standard Model.
This is yet another context where neutrino-neutrino interactions can play a crucial role in their flavor evolution.
\end{abstract}

\maketitle

%%%%%%%%%%%%%%%%%%%%%%%%%%%%%%%%%%%%%%%%%%%%%%%%%%%%%%%%%%%%%%%%%%%%%%%%%%%%%
\section{Introduction}
Core-collapse supernovae (CCSNe) and neutron star mergers (NSMs) commonly lead to a burst of thermal neutrinos in the MeV range, with a very active literature on the physics of their oscillations.  
%The IceCube discovery of high-energy neutrinos has opened a new avenue to investigate this astrophysical phenomenon and related neutrino physics. 
These enviroenments are considered as the central engine of not only supernovae but also gamma-ray bursts (GRBs), and other energetic or transrelativistic supernovae driven by outflows such as jets and winds (i.e., engine-driven supernovae). It has been suggested that GeV-TeV neutrinos can be produced in such environments if neutron-loaded outflows are launched from a black hole with an accretion disk and/or a newborn magnetar~\cite{Bahcall:2000sa,Meszaros:2000fs,Murase:2013hh,Kashiyama:2013ata,Murase:2013mpa}. Even TeV-PeV neutrinos can be generated inside the outflows through shock acceleration or magnetic reconnections~\cite{Meszaros:2001ms,Razzaque:2004yv,Ando:2005xi,Murase:2013ffa,Kimura:2018vvz}. 

\begin{figure} [tb!]%  figure placement: here, top, bottom, or page
 \centering
\begin{center}
\includegraphics*[width=.4\textwidth, trim= 0 10 0 50, clip]{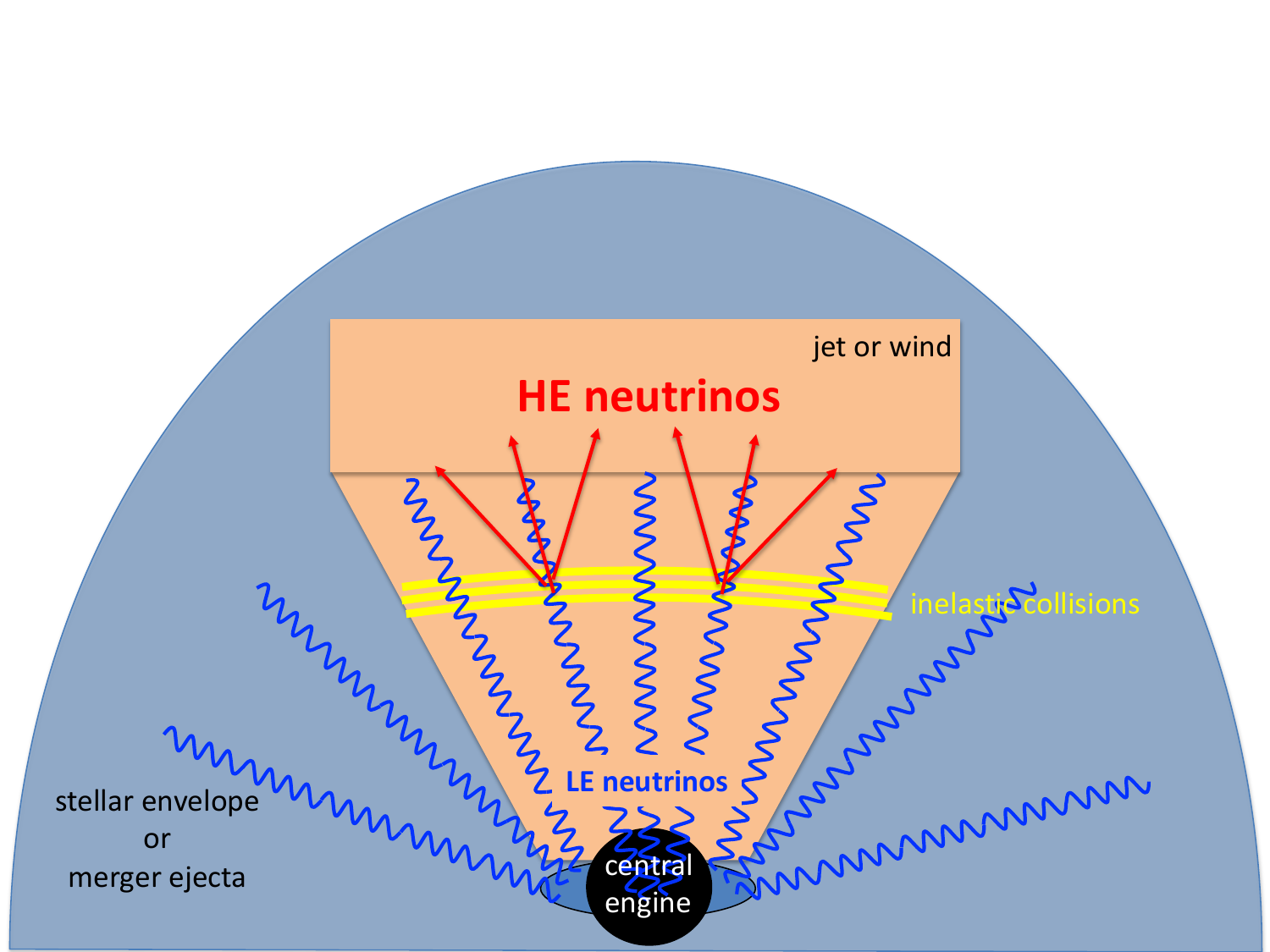}
\end{center}
\caption{Schematic picture of HE$\nu$ (GeV-PeV) production and their  interactions with LE$\nu$'s (MeV-GeV) from the central engine such as a black hole with an accretion disk or a newborn magnetar. HE$\nu$ production occurs at $R_{\rm diss}\gg R_{\rm eng}$, which may be beamed with the opening angle $\sim1/\Gamma$, with $R_{\rm diss}$, $R_{\rm {eng}}$, and $\Gamma$ being the dissipation radius, the engine radius, and the outflow Lorentz factor. Note that the opening angle of the LE$\nu$ beams is exaggerated for illustration purposes.
}
\label{fig:ref}
\end{figure}

The IceCube discovery of high-energy neutrinos (HE$\nu$'s) has opened a new avenue to investigate the physics of neutrino oscillations and related neutrino physics (see Refs.~\cite{Ackermann:2019cxh,Ackermann:2022rqc,Arguelles:2022xxa} and references therein). 
In this article, we investigate a novel effect caused by the interplay between the HE$\nu$'s produced in outflows and low-energy neutrinos (LE$\nu$'s) directly from the central engine (see Fig.~\ref{fig:ref} for the schematic picture). Indeed, the decohered LE$\nu$'s, which are in mass eigenstates, can provide a \emph{dominant} unusual background for the propagation of the HE ones. In particular, we show that the resulting neutrino self-interactions ($\nu$SI) leads to a very intriguing phenomenon, in which the HE$\nu$'s experience short-scale flavor oscillations in such a way that on average, they end up in the mass eigenstates.
This phenomenon is noncollective in spirit and differs remarkably from the well-known phenomenon of \textit{collective oscillations} of MeV neutrinos occurring in dense neutrino environments such as CCSNe and NSM 
remnants~\cite{Pastor:2002we,duan:2006an,duan:2006jv}.
 
%Such an interesting interplay is particularly plausible because the HE$\nu$ production can occur within the outflow duration or breakout time (that are longer than the light crossing time). This obviously implies that the HE$\nu$'s can propagate in the bath of the MeV neutrinos, as indicated schematically in Fig.~\ref{fig:ref}.  
%leading to neutrino oscillations induced by interactions between high-energy neutrinos and low-energy neutrinos

\section{HE neutrino interactions in jets or winds}
Various scenarios for HE$\nu$ production in GRBs, CCSNe and NSMs have been suggested. In this work, we are interested in the fate of HE$\nu$'s so we assume that they are produced at the dissipation radius $R_{\rm diss}\sim{10}^{8}-10^{10}~{\rm cm}$, which is much larger than the engine radius $R_{\rm eng}\sim10^{6}$~cm.  
Then an interesting interplay is particularly plausible when the HE$\nu$ production occurs within the duration of LE$\nu$'s and outflow breakout time (that are longer than the light crossing time). Note that as seen below the effect on neutrino oscillation is largely model independent as long as $R_{\rm diss}$ is so small that LE$\nu$'s govern the neutrino potential.  

Although the proposed mechanism works in pretty general setups, for illustrative purposes, we consider models of GeV-TeV neutrinos. Quasithermal neutrinos can naturally be produced in the GeV-TeV range through inelastic neutron-proton collisions when neutrons decouple from protons or neutron-loaded outflows make collisions with the surrounding environment~\cite{Bahcall:2000sa,Meszaros:2000fs,Murase:2013hh,Murase:2013mpa}, and higher-energy nonthermal neutrinos may also be produced through neutron-proton-converter acceleration~\cite{Murase:2013hh,Kashiyama:2013ata}. For these neutrinos, the dissipation may occur at $R_{\rm diss}\sim{10}^{8}-10^{10}$~cm~\cite{Bahcall:2000sa,Murase:2013mpa}.
Protons could further be accelerated to higher energies via shock acceleration or magnetic reconnections, and nonthermal TeV neutrinos can be efficiently produced via inelastic $pp$ and/or $p\gamma$ interactions~\cite{Meszaros:2001ms,Razzaque:2004yv,Ando:2005xi,Murase:2013ffa,Kimura:2018vvz}. These neutrinos are associated with the dissipation at internal, collimation, and termination shocks~\cite{Murase:2013ffa,Meszaros:2001ms,Murase:2013mpa,Kashiyama:2013qet,Murase:2017pfe}. For example, the internal dissipation radius is estimated to be $R_{\rm diss}\approx2\Gamma^2 c \delta t\sim 6\times{10}^{8}~{\rm cm}~{(\Gamma/3)}^2~(\delta t/1~{\rm ms})$, where $\delta t$ is the variability time. 

The number density of LE$\nu$'s at $R_{\rm diss}$ (in the engine frame) is 
\begin{eqnarray}
n_{\mathrm{LE}\nu}=\frac{L_{\nu_e}}{4\pi R_{\rm diss}^2 c \langle E_\nu\rangle}
&\simeq&1.7\times10^{27}~{\rm cm}^{-3}~{\left(\frac{L_{\nu_e}}{10^{52}~{\rm erg}~{\rm s}^{-1}}\right)}\nonumber\\
&\times&{\left(\frac{R_{\rm diss}}{10^{9}~{\rm cm}}\right)}^{-2}{\left(\frac{\langle E_\nu\rangle}{10~{\rm MeV}}\right)}^{-1},
\end{eqnarray}
which can be much larger than the expected number density of HE$\nu$'s therein, $n_{\mathrm{HE}\nu} \lesssim 10^{24}\ \rm{cm}^{-3}$ at $10^9$~cm.
Here $L_{\nu_e}$ and $\langle E_\nu\rangle$ are the electron neutrino luminosity and average energy, respectively. 
In addition, the electron number density in the outflow is 
\begin{eqnarray}
n_e \approx \frac{\Gamma L}{4\pi R_{\rm diss}^2 \Gamma^2 m_pc^3}&\simeq&5.9\times10^{23}~{\rm cm}^{-3}~{\left(\frac{L}{10^{52}~{\rm erg}~{\rm s}^{-1}}\right)}\nonumber\\
&\times&{\left(\frac{R_{\rm diss}}{10^{9}~{\rm cm}}\right)}^{-2}{\left(\frac{\Gamma}{30}\right)}^{-1} \ll n_{\mathrm{LE}\nu}.
\end{eqnarray}
%where $\Gamma$ is the outflow Lorentz factor.

Unlike the flavor evolution of the LE$\nu$'s which is dominated by the mass Hamiltonian at such neutrino number densities,
%(occurring on scales
%determined by  the neutrino vacuum frequency, $\omega = \Delta m^2/2E$,
%which is $ \sim$ 1 km$^{-1}$ for a $E =10$ MeV neutrino and atmospheric mass difference), 
the evolution of HE$\nu$'s can be dominated by their coherent scattering with the bath of the LE$\nu$'s.
This simply comes from the fact that for the HE$\nu$'s, the strength of $\nu$SI (see Eq.~(\ref{Hnunu})), 
\begin{equation}
%\mu=\sqrt2 G_\mathrm{F} n_{\nu_e} \xi \simeq 6.4\times{10}^{-3}~\xi_{-2}~\rm{km}^{-1} \bigg(\frac{n_{\nu_e}}{10^{27}\ \rm{cm}^3}\bigg) ,
\mu \approx \sqrt2 G_\mathrm{F} n_{\nu} \hbar^2c^2\ \xi \simeq 6.4 \times 10^{-6}\ {\rm cm}^{-1} \bigg(\frac{n_{\nu}}{10^{27}\ {\rm cm}^3}\bigg)\ \xi,
\end{equation}
can be much larger than their vacuum wavelength, $\omega\approx\Delta m_{\rm{atm}}^2c^3/(2\hbar E_\nu) \simeq 6\times10^{-10}\ \rm{cm}^{-1}\ \big(100\ \mathrm{GeV}/\it{E}_\nu\big)$,
with $ G_\mathrm{F}$ being the Fermi constant.
In the above equation $\xi = 1-\cos\Theta$, where $\Theta$ is the opening angle of the neutrino beams,
which is determined here mainly by the opening angle of HE$\nu$'s.  
Note that  as soon as the parameter $\mu$ is known, $\xi$ and $\nu$ do not provide any more relevant information.
%\km{If HE neutrinos are produced by relativistic flows with $\Gamma\gg1$, we have $\xi\approx\Theta^2/2\sim 1/(2\Gamma^2)\simeq 0.05~\Gamma_1^{-2}$.}
For relativistic flows with $\Gamma\sim2-100$, one has $\xi\approx\Theta^2/2\sim 1/(2\Gamma^2)$. 
Note that the optical depth to incoherent neutrino scatterings is so small that the electron-positron pair production is negligible. 
Moreover, given the fact that the number density of LE$\nu$'s is much larger than that of HE$\nu$'s, one can assume that $n_{\nu}$ is here  exclusively determined by the LE$\nu$'s.
 
Although the number density of the LE$\nu$'s within the zones of interest is expected to be too small to allow for the $\nu$SI Hamiltonian to compete with or dominate their vacuum Hamiltonian, the evolution of HE$\nu$'s is almost completely governed by the interaction term for appropriate 
LE$\nu$ number densities ($\omega_{\mathrm{HE}\nu} \ll \mu \lesssim \omega_{\mathrm{LE}\nu}$).

\section{Two-Beam Model}
In order to demonstrate how the flavor content of HE$\nu$'s is impacted by their propagation in the bath of the LE$\nu$'s, we study neutrino flavor conversions in a one-dimensional two-beam model, which 
%is basically the one-dimensional version of similar to the one studied first in Ref.~\cite{duan:2014gfa}, but
consists of \textit{two} energy bins, and a \textit{three-flavor} neutrino gas with two angular beams.  
The neutrino energies are taken to be $E_\nu = 10$~MeV and  $100$~GeV for the bins representing the LE$\nu$'s and the HE$\nu$'s, respectively, unless otherwise stated.
%The  $\nu_e$ and $\bar{\nu}_e$ beams are assumed to be emitted with $\mathbf{v}_{\pm} = (\pm u,0,v_z)$, where $u = \sqrt{1-v_z^2}$ with $v_z=1/2$ (corresponding to an opening angle of $2\pi/3$ between the two beams).
Thus in brief, our model consists of two angle beams each including neutrinos and antineutrinos with two energies representing high- and low-energy neutrinos.
We also assume that the neutrino density is constant within the bath of LE$\nu$'s.
 
In order to study the flavor evolution of neutrinos in our model, we solve the Liouville-von Neumann equation  for the neutrino density matrix, $\varrho$ ($c=\hbar=1$)
\cite{Sigl:1992fn}%,Strack:2005ux,Cardall:2007zw,Volpe:2013jgr, Vlasenko:2013fja}
\begin{equation}
i d_t
\varrho_{\mathbf{p}} = \left[
  \frac{\mathsf{U}\mathsf{M}^2 \mathsf{U}^{\dagger}}{2E_\nu} + \mathsf{H}_{\mathrm{m}} +
  \mathsf{H}_{\nu \nu, \mathbf{p}} ,
  \varrho_{\mathbf{p}}\right],
\label{Eq:EOM}
\end{equation} 
with
\begin{equation}\label{Hnunu}
  \mathsf{H}_{\nu \nu, \mathbf{p}} = \sqrt2 G_{\mathrm{F}}
  \int\!  \frac{\mathrm{d}^3p'}{(2\pi)^3}
  ( 1- \mathbf{v} \cdot \mathbf{v}')
  (\varrho_{\mathbf{p}'} - \bar\varrho_{\mathbf{p}'}),
\end{equation}
being the neutrino potential stemming from the neutrino-neutrino forward scattering~\cite{Fuller:1987aa,Pantaleone:1992xh,Notzold:1988kx}.
Here $\mathbf{p}$ is the neutrino momentum, $E_\nu=|\mathbf{p}|$, $\mathbf{v} = \mathbf{p}/E_\nu$, and $\mathsf{M}^2$ are the energy, velocity, and mass-square matrix of the neutrino, respectively, and $\mathsf{U}$ is the Pontecorvo–Maki–Nakagawa–Sakata matrix.
Moreover, $\mathsf{H}_{\mathrm{m}}$ is the contribution from the matter term which is proportional to matter (electron) density~\cite{Wolfenstein:1977ue,Mikheev:1986gs}, which is ignored in our calculations due to the relatively small matter density inside the outflow. Hence,  there are only two nonzero terms  in $\mathsf{H}$ (vacuum and $\nu$SI)
which are both diagonal in the mass basis and constant (see below), but with different eigenvalues.
 
As mentioned above, $\mathsf{H}_{\nu\nu}$ is almost exclusively determined by LE$\nu$'s here due to their much larger number densities. In this study, we assume that LE$\nu$'s are in mass eigenstates because they are expected to be already decohered within the zones of interests, which are very far from their emission region (with a typical coherence length of a few $10^6$~cm for the atmospheric mass difference)~\cite{Beuthe:2001rc, Akhmedov:2009rb,Kersten:2015kio}. 
In addition, the HE$\nu$'s do not significantly disturb the flavor state of the LE$\nu$ bath due to their much smaller number densities.
%\textcolor{blue}{leaving us with the integral $[\rho_\mathbf{p},\rho_\mathbf{p'}]$ for the LE$\nu$'s (Here, I am not sure how many angular nodes we're using. My understanding is that there is no spread in the beam, which would make the commutator 0. Please confirm.)}.                                          
This implies that the LE$\nu$'s do not evolve since their evolution is dominated by the mass term and they are already in the mass eigenstates. Consequently, $\mathsf{H}_{\nu\nu}$ remains approximately constant.  
(We will later discuss how the neutrino flavor evolution changes once the oscillations of LE$\nu$'s are taken into account.)
On the other hand, the HE$\nu$'s find themselves in the bath of the LE$\nu$'s as soon as they are produced, and start flavor conversions on relatively short scales due to their interactions with the LE$\nu$'s. It should be kept in mind  that the coherence length of HE$\nu$'s
%,$L_{\mathrm{coh}} \sim \sigma_x E^2/\Delta m^2$, 
are expected to be much longer than those of the LE ones~\cite{Farzan:2008eg}. 
%\km{If the coherence is not lost, the flavor result should be modified from Eq.8 due to the vacuum oscillation??}
%assuming $\sigma_x \propto 1/E$ due to the 
%uncertainty principle, and with $\sigma_x$ being the production size of the neutrino wave packet.

\section{Results}
\begin{figure} [t!]%  figure placement: here, top, bottom, or page
\centering
\begin{center}
\includegraphics*[width=.42\textwidth, trim= 0 10 0 10, clip]{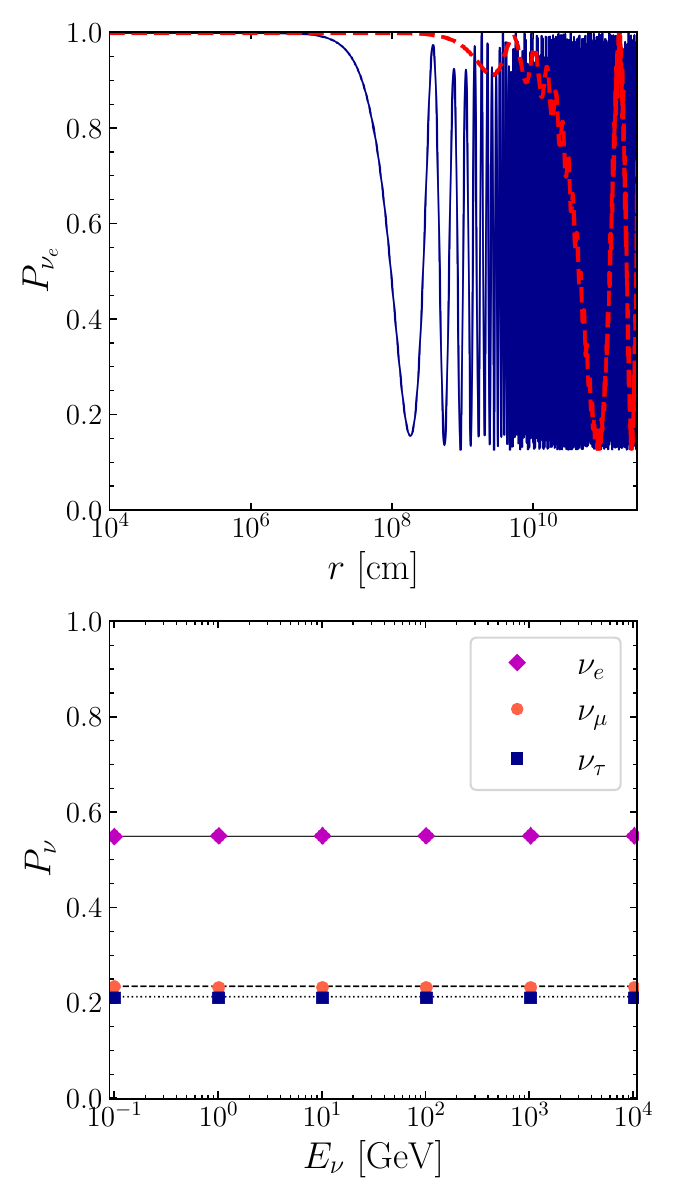}
\end{center}
\caption{Upper panel:  
Survival probability of HE$\nu_e$'s propagating  in vacuum (dashed red curve), and in the bath of the LE$\nu$'s (solid blue curve). HE$\nu$'s can experience flavor conversions on scales much shorter than those expected in vacuum.
Here, for illustrative purposes we have assumed $\mu=10^{-7}$~cm$^{-1}$, $E_{\rm{LE}\nu} = 10$~MeV and $E_{\rm{HE}\nu} = 100$~GeV. For the LE$\nu$ bath, $n_{\bar{\nu}_e}/n_{\nu_e}=1.3$ is also fixed, although the survival probabilities are independent of $n_{\bar{\nu}_e}/n_{\nu_e}$  as long as it is not too close to 1. 
Lower panel: The survival probabilities  as a function of the HE$\nu$ energy, where the diamonds, points, and squares are the survival probabilities of $\nu_e$, $\nu_\mu$, and $\nu_\tau$ obtained from the simulations, respectively, and the black lines are the corresponding analytical solutions. We assume an  initial flavor ratio of $\nu_e:\nu_\mu:\nu_\tau=1:0:0$.
We here set $\theta_{12} = 33.6^{\circ}, \theta_{23}=47.2^{\circ}$, $\theta_{13}=8.5^{\circ}$ and $\delta_{\rm{CP}}=0$.
%The expected flavor ratio does not show any energy dependence. 
Antineutrinos behave exactly in the same manner.
}
\label{fig:nu}
\end{figure}

In the upper panel of Fig.~\ref{fig:nu}, we show the survival probabilities of HE$\nu_e$'s propagating in vacuum (dashed curve), and in the bath of the LE$\nu$'s (solid curve). As can be clearly seen, the oscillation scales of HE$\nu$'s can change by orders of magnitude when coherent scatterings with LE$\nu$'s are taken into account. 
As a matter of fact, the  oscillation scale of HE$\nu$'s in a bath of the LE ones is determined by the number density of the LE$\nu$'s, namely $l_{\rm{osc}}\sim |\mathsf{H}_{\nu\nu}|^{-1} \sim\mu^{-1}$ ($\sim10^5$~cm for this simulation). 
This scaling behavior can be immediately deduced from Eq.~(\ref{Eq:EOM}) given the fact that for the HE$\nu$'s, the dominant contribution to the Hamiltonian comes from coherent scatterings with LE$\nu$'s as long as $ \omega_{\rm{HE\nu}} \ll \mu$ (we indeed observe this behaviour for $10~\omega_{\rm{HE\nu}} \lesssim \mu$).
Note that here the only relevant physical parameter is $\mu/ \omega_{\rm{HE\nu}}$, therefore  one can play with $\mu$ in Fig.~\ref{fig:nu} as long as this ratio is constant (provided that $r$ is also appropriately rescaled for the upper panel).
%\km{We also note that the LE$\nu$'s may have $n_{\bar{\nu}_e}/n_{\nu_e}>1$ (e.g., Refs.~\cite{Liu:2015prx,Uribe:2019cpq,Just:2022flt}).}

The fact that HE$\nu$'s oscillate on scales $\sim \mu^{-1}$ might remind an astute reader of the phenomenon of \textit{fast} flavor conversions occurring for MeV neutrinos in dense neutrino media~\cite{Sawyer:2005jk, Sawyer:2015dsa}. This similarity becomes more obvious once one notes that 
the oscillations of HE$\nu$'s can even occur when $\omega_{\rm{HE\nu}} = 0$.
However and in spite of this resemblance, it should be kept in mind that these two phenomena completely differ in spirit and \textit{have nothing to do with each other}. Although for the occurrence of fast conversions certain criteria need to be fulfilled~\cite{Morinaga:2021vmc}, the short-scale conversions of HE$\nu$'s in a bath of the LE ones is a generic phenomenon provided that there are two populations of neutrinos of which one is dominant.

Flavor conversions of HE$\nu$'s induced by the static bath of the LE$\nu$'s is also distinct from the phenomenon of ordinary collective oscillations in dense neutrino media. While the latter is a nonlinear phenomenon with a high level of coupling, the former is a linear phenomenon where $\mathsf{H}_{\nu\nu}$ solely provides a constant background for the flavor evolution of HE$\nu$'s. 
This implies that such flavor conversions of HE$\nu$'s is a noncollective phenomenon. Also note that the relevant $n_\nu$'s can be many orders of magnitude smaller than the values for which collective oscillations of MeV neutrinos are expected, due to the much smaller $\omega_{\rm{HE\nu}}$.

%One can clearly see that the vacuum term also plays a role on larger scales ($\sim 10^{10}$~cm) and leads to another oscillatory behavior on top of the one on shorter scales. This, of course, should be expected since this is the scale on which the vacuum term becomes relevant. 
%brought about by coherent scatterings with LE$\nu$'s.
%Despite this, the relevant scale of these oscillations is too long and consequently, one might expect  full decoherence of HE$\nu$'s by then. 

In order to see how the short-scale flavor conversions of HE$\nu$'s changes the expected $\nu_e :\nu_\mu:\nu_\tau$ ratio on earth, one can average the survival probabilities over a few oscillations.
%i.e. over a few $L_{\rm{osc}}$'s. 
As indicated in our upcoming work~\cite{unpublished}, such an averaging process in our two-beam model corresponds to averaging over the neutrino angular distribution
in a more-realistic, multiangle neutrino gas. The average survival probabilities then reach a steady 
state %~\footnote{For our two-beam model, it means it does not matter which cycles are selected for averaging.} 
which does not depend on the details of the simulation (apart from the neutrino mixing parameters as discussed in the following), shown in the lower panel of Fig.~\ref{fig:nu}. 
%Surprisingly, the  expected flavor ratio $\nu_e : \nu_\mu : \nu_\tau$  on Earth
%is independent of the neutrino energy. Moreover, the initial flavor state of the neutrino, i.e. whether it was 
%an electron, a muon, or a tau neutrino, or even antineutrino,   does not matter here. 
%The average survival probabilities of  HE$\nu$'s neutrinos can be indeed understood analytically.
%It turns out that one can analytically predict the average survival probabilities of  HE$\nu$'s neutrinos. 
This behaviour can be understood analytically as follows.
% This can be seen by noting that 
%$\langle P_\nu\rangle$ reaches a stationary sate provided that,
%\begin{equation}
%[\langle\rho\rangle, \mathsf{H}_{\nu\nu}] \simeq 0.
%\end{equation}
%This equality is always satisfied if $\langle\rho\rangle \propto \mathsf{H}_{\nu\nu}$. 
In the mass basis, the $\nu$SI Hamiltonian is diagonal
with its $kk$-th component being $h_k \propto \sum_\alpha |U_{\alpha k}|^2 (\rho_{\alpha \alpha} - \bar\rho_{\alpha \alpha})$, where $\overset{\textbf{\fontsize{2pt}{2pt}\selectfont(--)}}{\rho}_{\alpha \alpha}$ are the initial (anti)neutrino occupation numbers in flavor $\alpha$. This comes from the fact that
$\mathsf{H}_{\nu\nu}$ is nearly determined only by the LE$\nu$'s  which are 
%already decohered and consequently, are 
in the mass eigenstates.    
%Using this relation and considering the fact that  
%\begin{equation}
%\mathsf{H}^{\rm{flavor\ basis}}_{\nu\nu} = \mathsf{U}\ \mathsf{H}^{\rm{mass\ basis}}_{\nu\nu}\ \mathsf{U}^{\dagger}, \nonumber
%\end{equation}
%one can then find out equilibrium $\langle\rho\rangle$ apart from a multiplication factor.  
Then the HE$\nu$ density matrix in the mass basis, $\tilde\rho$, evolves as,
\begin{equation}\label{eq:rho_mass}
\tilde\rho_{ij}(t) = \tilde\rho_{ij}(0) \ e^{-\mathrm{i}(h_i-h_j)t},
\end{equation}
implying that the averaged flavor ratio, $\nu_e:\nu_\mu:\nu_\tau$, can be written as,
%\lipsum[1]
%\begin{widetext}
%\begin{widetext}
\begin{equation}
|U_{ek}|^2 |U_{\alpha k}|^2 f_\alpha : |U_{\mu k}|^2 |U_{\alpha k}|^2 f_\alpha : |U_{\tau k}|^2 |U_{\alpha k}|^2 f_\alpha,
\label{ratio_}
\end{equation}
% \end{widetext}
 %\end{widetext}
%\lipsum[1]
where there is a summation over $\alpha$ and $k$, and $f_e:f_\mu:f_\tau$ is the initial flavor ratio at the production region. The black lines in the lower panel of Fig.~\ref{fig:nu} indicate the analytical flavor ratios in Eq.~(\ref{ratio_}), which show a perfect agreement with the numerical results.

Note that the average density matrix in the flavor basis is  equal to the one expected after the neutrino decoherence, and is also independent of the neutrino energy.
This behaviour indeed results from the fact that the HE$\nu$'s oscillate very quickly about $\mathsf{H}_{\nu\nu}$ and consequently, they end up  in the mass eigenstates (on average). 
Hence, in summary, short-scale conversions of HE$\nu$'s induced by the ambient gas of the LE$\nu$'s lead to their decoherence on scales which can be shorter than their natural decoherence length~\cite{Farzan:2008eg} by \emph{many orders of magnitude}.

\begin{figure} [tb!]%  figure placement: here, top, bottom, or page
\centering
\begin{center}
\includegraphics*[width=.43\textwidth, trim= 35 220 30 230, clip]{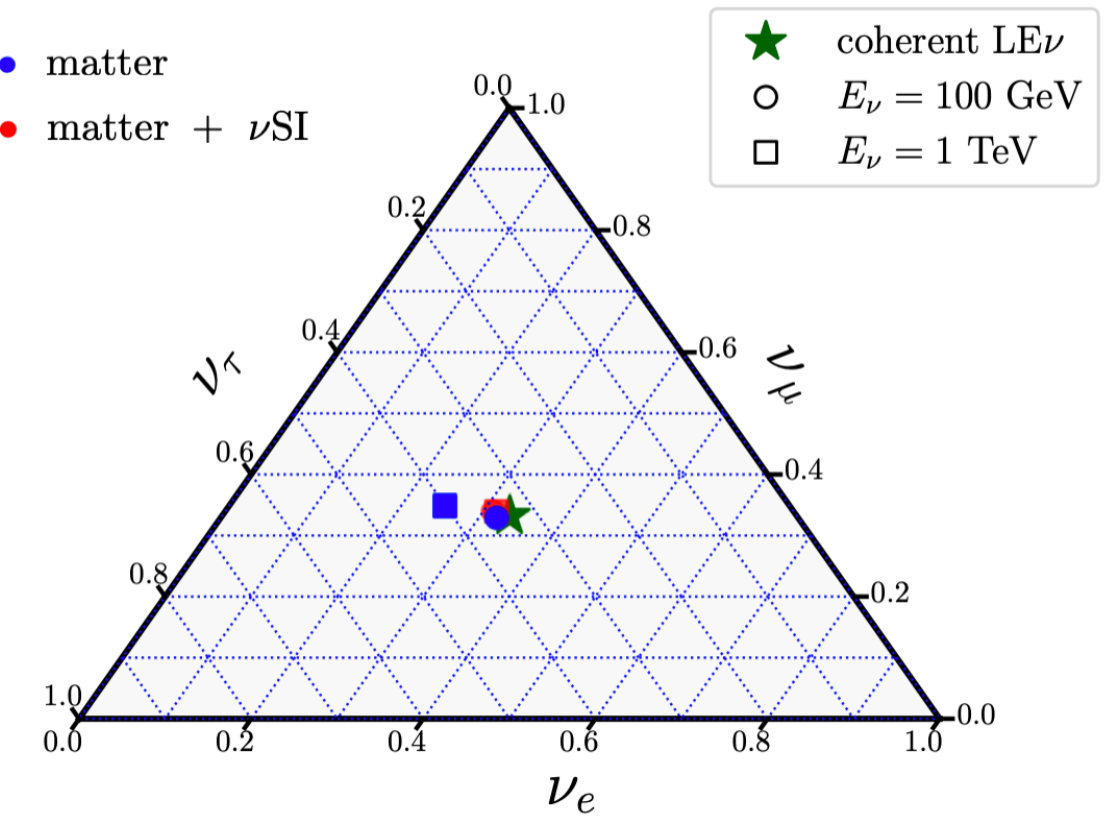}
\end{center}
\caption{The expected $\nu_e:\nu_\mu:\nu_\tau$ ratio on Earth in the absence and presence of $\nu$SI for different neutrino energies. Note that the matter effect is included in both cases, assuming the density profile of a blue supergiant, and the ratio with $\nu$SI is very close to the total flavor equipartition. 
In addition, the green star indicates the total flavor equipartition expected from the propagation of HE$\nu$'s in a bath of \textit{oscillating} LE$\nu$'s, as discussed in the 
text. Note that apart from the matter only case for  the 1 TeV neutrinos, the other ones are almost on top of each other.
It is also illuminating to keep in mind that  the final flavor states are specific to the initial flavor composition of 1:2:0 and can vary under different circumstances.}
\label{fig:MSW}
\end{figure}

Although such short-scale oscillations and the resulting decoherence of HE$\nu$'s is an interesting phenomenon by itself and could in principle impact the physics of their propagation by modifying their flavor ratio at the source, %\textcolor{blue}{\sout{\km{refs}}},
we here discuss a few important cases in which the induced conversions of HE$\nu$'s can be observable on Earth.

Once HE$\nu$'s leave the LE$\nu$ bath, they should propagate in the dense ejecta, where $n_e\gg n_{\nu}$~\cite{Mena:2006eq,Sahu:2010ap,Razzaque:2009kq, Xiao:2015gea,Carpio:2020app}. 
In particular, for engine-driven transients, the HE$\nu$ production region is surrounded with the stellar or merger ejecta, whose density is much larger than that in the production region. 
This means that in solving Eq.~(\ref{Eq:EOM}) for the neutrino propagation in this region, we  ignore the $\mathsf{H}_{\nu \nu}$ term. 
In order to account for the decoherence experienced by neutrinos in the LE$\nu$ bath, we start with an initial density matrix which is a time average 
of the one in Eq.~(\ref{eq:rho_mass}). In addition, for the $\mathsf{H}_{\mathrm{m}}$ term we consider 
 %In this region, we expect that the flavor ratio is further affected by the matter effect, where we use 
a blue supergiant matter profile from Ref.~\cite{Woosley:1995ip}, as an example ($30~M_\odot$ BSG in Ref.~\cite{Carpio:2020app}).
Needless to say, the short-scale conversions of HE$\nu$'s will impact the outcome of the matter effect and correspondingly, their flavor ratio on Earth.
This is illustrated clearly in Fig.~\ref{fig:MSW} for a case with the initial flavor ratio $\nu_e:\nu_\mu:\nu_\tau=1:2:0$ in the normal mass ordering.
This is particularly interesting considering the energy-independent nature of the LE$\nu$-induced short-scale oscillations of HE$\nu$'s.
Although the pure matter effect shows a clear sign of energy dependence here~\cite{Carpio:2020app}, it is almost independent of the neutrino energy in the presence of $\nu$SI. This is even the case at high energies where the muon damping is expected to occur in such dense environments~\cite{Kashti:2005qa,Karmakar:2020yzn}.
This can provide one with a new observable indication of neutrino flavor mixing caused by $\nu$SI. 

So far we have assumed that the bath of the LE$\nu$'s is completely decohered. However, the situation could be different. On the one hand, since $R_{\mathrm{diss}}$ can be as low as $10^8$~cm, the solar-mass channel LE$\nu$'s could be still in phase since $l_{\mathrm{coh}, \odot} \sim \sigma_x E_\nu^2/\Delta m_{\rm{sol}}^2 \gtrsim 10^{8}$~cm.  
Moreover, due to the possibility for the existence of $\mu \gtrsim 10^{-4} \ \rm {cm}^{-1}$, the decoherence of LE$\nu$'s \textit{might} be suppressed~\cite{Akhmedov:2017mcc} in the atmospheric channel as well and LE$\nu$'s can experience a sort of (partial) collective oscillations in the production region of HE ones. Such erratic conversions of LE$\nu$'s will lead to the total flavor equipartition of HE$\nu$'s regardless of their initial flavor content, as indicated by the green star in Fig.~\ref{fig:MSW}. This phenomenon will be discussed in more details in our upcoming work~\cite{unpublished}\footnote{The only difference of this case with the results shown in Fig.~\ref{fig:nu} is that
 we here allow for flavor oscillations of the LE$\nu$ gas rather than fix it to be in the mass state.}.

\begin{figure} [bth!]%  figure placement: here, top, bottom, or page
\centering
\begin{center}
\includegraphics*[width=.43\textwidth, trim=  60 30 50 30, clip]{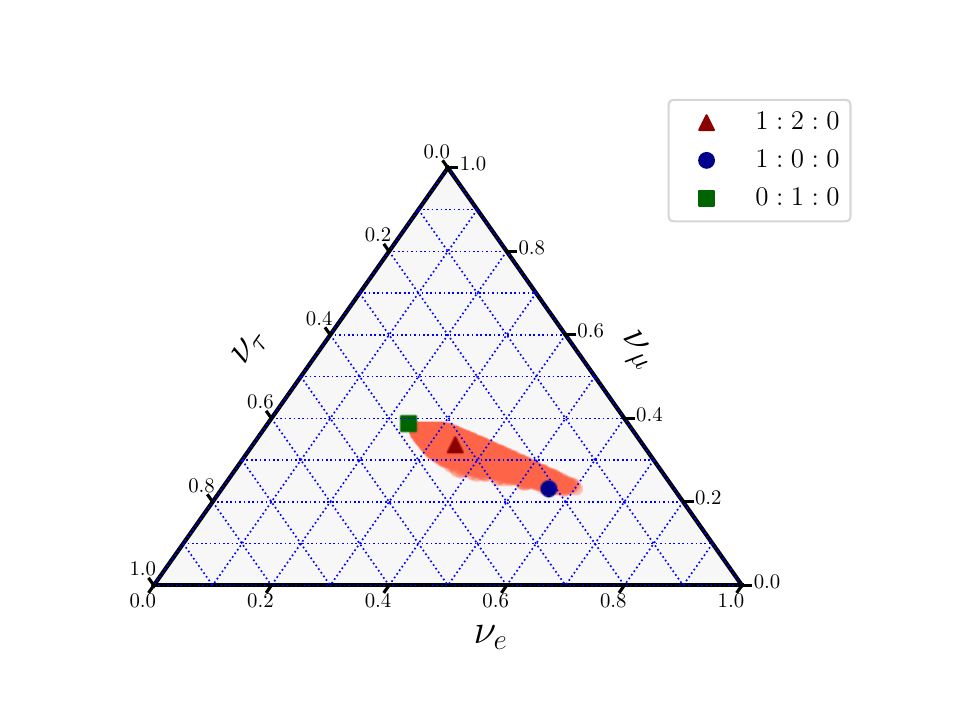}
\end{center}
\caption{
The expected $\nu_e:\nu_\mu:\nu_\tau$ ratio after HE$\nu$'s escape their production region, in the presence of $\nu$NSSI. The triangle, circle, and square indicate the ratio in SM while the red region shows how the ratio changes
in the presence of $\nu$NSSI for the $1:2:0$ case. 
Here the red region is created by choosing a large set of  randomly populated $\hat{G}^{\alpha\beta}$
assuming that $|\hat{G}^{\alpha\beta}|<1$ (for $\alpha \neq \beta$).
Via coherent $\nu$SI, the final HE$\nu$ flavor ratio is very sensitive to the $\nu$NSSI.}
\label{fig:G}
\end{figure}

In some of the beyond Standard Model (SM) theories of particle physics, neutrinos can experience neutrino nonstandard self-interactions ($\nu$NSSI)~\cite{Bialynicka-Birula:1964ddi,Bardin:1970wq}. Such $\nu$NSSI modify Eq.~(\ref{Hnunu}) to~\cite{sigl1993general,Blennow:2008er,Das:2017iuj}
\begin{align}\label{eq:G}
\mathsf{H}_{\nu \nu, \mathbf{p}} = \sqrt2 G_{\mathrm{F}}
\int\!  \frac{\mathrm{d}^3p'}{(2\pi)^3} &( 1- \mathbf{v} \cdot \mathbf{v}')
\{\hat{G}(\varrho_{\mathbf{p}'} - \bar\varrho_{\mathbf{p}'})\hat{G}\nonumber \\
&+\hat{G}\ \mathrm{Tr}[(\varrho_{\mathbf{p}'} - \bar\varrho_{\mathbf{p}'})\hat{G}]   \},
\end{align}
where $\hat{G}$ contains information about $\nu$NSSI ($\hat{G}=\mathbb{1}$ in SM). 
For example, in the vector mediator scenario, we may have $\mathcal{L}_{\mathrm{eff}} \supset G_{\mathrm{F}} [\hat{G}^{\alpha\beta} \bar\nu_\alpha \gamma^{\mu} \nu_\beta] [\hat{G}^{\xi\eta} \bar\nu_\xi \gamma_{\mu} \nu_\eta]$, and its $\nu$NSSI components are related to the vector mediator mass $m_V$ and the coupling strength $g$ by $|\hat{G}^{\alpha\beta}| \propto g^2/m_V^2$. 
The current constraints on $\nu$NSSI are model dependent and strong for the mediator mass below MeV energies. For heavier mediators, the constraints from the early universe are rather weak, e.g., $|\hat{G}^{\alpha\beta}| \lesssim 10 ^{7}$~\cite{Archidiacono:2013dua}, although laboratory constraints can be stronger in the limited parameter space~\cite{Berryman:2022hds}. 
%$|\hat{G}^{\alpha\beta}| \lesssim 10$, that can be ruled out by the Z-decay~\cite{bilenky1994bounding}).
It has been suggested that spectral modulations and time delays of HE$\nu$'s enable us to study the unexplored parameter space of $\nu$NSSI~\cite{Ioka:2014kca,Ng:2014pca,Ibe:2014pja,2014arXiv1408.3799B,DiFranzo:2015qea,Shoemaker:2015qul}.
We point out that coherent $\nu$SI-induced oscillations of HE$\nu$'s can be used as a novel probe of $\nu$NSSI. 
This is illustrated in Fig.~\ref{fig:G} where the red region shows the impact of $\nu$NSSI.
The flavor content is expected to have observable sensitivity to $\nu$NSSI, 
i.e., a  $\sim10$\% change of flavor ratio is caused by $|\hat{G}^{\alpha\beta}|\sim 0.1$. This means that one could probe 
such weak couplings with this effect.
%and we could probe the coupling as weak as $|\hat{G}^{\alpha\beta}|\sim 0.1$ corresponding to the change of flavor ratios by $\sim10$\%. 

\section{Conclusion}
We have brought to light a novel phenomenon, in which a class of high-energy cosmic neutrino emission can experience flavor conversions induced by the copious LE$\nu$ background, on scales much shorter than their intrinsic vacuum oscillation wavelengths. Unlike the celebrated phenomenon of collective oscillations of MeV neutrinos in a dense neutrino medium, the unearthed flavor conversions of high-energy cosmic neutrinos is a noncollective phenomenon in spirit. 

This intriguing phenomenon can occur when HE$\nu$'s from relativistic outflows launched at the core-collapse of massive stars or at the mergers propagate in the bath of the already-decohered lower energy neutrinos from the central engine.
Despite the small number density of HE$\nu$'s which can be insufficient to result in their own collective oscillations, their presence can lead to short-scale conversions of HE$\nu$'s on scales determined by the density of LE$\nu$'s. The background-induced conversions of HE$\nu$'s change their flavor content in an energy-independent manner and takes the HE$\nu$ gas to a state which is diagonal in the mass basis. This way they cause an induced decoherence of HE$\nu$'s on scales which are many orders of magnitude shorter than their natural decoherence lengths. Such a modification of the HE$\nu$'s at source can impact the physics of the phenomena occurring during their propagation, such as neutrino decay, scattering, etc. We also point out a few possibilities where such short-scale induced decoherence can directly impact the flavor ratio of HE$\nu$ on Earth, including the matter effect of HE$\nu$'s, the $\nu$NSSI, and the total flavor equipartition due to an oscillating ambient LE$\nu$ gas.

%Besides MeV neutrinos in CCSNe, NSM, and in the early universe, our work uncovers yet another context where neutrino-neutrino interaction can play a crucial role in the physics of neutrinos flavor evolution. 
Our study provides the first step toward understanding
this intriguing phenomenon and further exploration is needed to better understand its implications. 
This is yet another context where neutrino-neutrino interactions can play a crucial role in their flavor evolution, and also motivates further investigations into \textit{multimessenger} high-energy emission from GRBs, CCSNe and NSMs.

%%%%%%%%%%%%%%%%%%%%%%%%%%%%%%%%%%%%%%%%%%%%%%%%%%%%%%%%%%%%%%%%%%%%%%%%%%%%%
% Place all of the references you used to write this paper in a file
% with the same name as following the \bibliography command
%%%%%%%%%%%%%%%%%%%%%%%%%%%%%%%%%%%%%%%%%%%%%%%%%%%%%%%%%%%%%%%%%%%%%%%%%%%%%

\section*{Acknowledgments}
We are  grateful to Georg Raffelt and Huaiyu Duan for very insightful discussions and their comments on the manuscript. 
S.A. acknowledges support by
 the German Research Foundation (DFG) through
the Collaborative Research Centre  ``Neutrinos and Dark Matter in Astro-
and Particle Physics (NDM),'' Grant SFB-1258, and under Germany’s
Excellence Strategy through the Cluster of Excellence ORIGINS
EXC-2094-390783311.
This work of K.M. is supported by the NSF Grant No.~AST-1908689, No.~AST-2108466 and No.~AST-2108467, and KAKENHI No.~20H01901 and No.~20H05852.
 %\begin{acknowledgments}

%\end{acknowledgments}

%\section*{References}
%\blindtext \cite{article-minimal}

\bibliographystyle{elsarticle-num}%apsrev4-1}
\bibliography{Biblio}
% \bibliography{ref}

%%%%%%%%%%%%%%%%%%%%%%%%%%%%%%%%%%%%%%%%%%%%%%%%%%%%%%%%%%%%%%%%%%%%%%%%%%%%%
\clearpage
%\appendix

\end{document}